%% file: main.tex
\documentclass[a4paper]{llncs}

\usepackage{amssymb}
\usepackage{graphicx}
\usepackage{wrapfig} 
\usepackage{enumerate}
\usepackage[shortlabels]{enumitem}
\usepackage{todonotes}
\usepackage{pifont}

\usepackage{enumitem}

\usepackage{tgtermes} 
\usepackage{caption}
\usepackage{subcaption}
\usepackage{url}
\urldef{\mailsa}\path|{johan.linaker,bjorn.regnell}@cs.lth.se|
\urldef{\mailsb}\path| {patrick.rempel,patrick.maeder}@tu-ilmenau.de|
\newcommand{\keywords}[1]{\par\addvspace\baselineskip
\noindent\keywordname\enspace\ignorespaces#1}

\begin{document}

\mainmatter  

\title{How Firms Adapt and Interact in Open Source Ecosystems: Analyzing Stakeholder Influence and Collaboration Patterns}

\titlerunning{How Firms Adapt and Interact in Open Source Ecosystems}

%
%
\author{Johan Lin{\aa}ker\inst{1}, Patrick Rempel\inst{2}, Bj{\"o}rn Regnell\inst{1}, and Patrick M{\"a}der\inst{2}}
\authorrunning{Johan Lin{\aa}ker, Patrick Rempel, Bj{\"o}rn Regnell, and Patrick M{\"a}der}

\institute{Lund University, Sweden\\
\mailsa
\and Technische Universit{\"a}t Ilmenau, Germany\\
\mailsb}

%
%

\toctitle{Lecture Notes in Computer Science}
\tocauthor{Authors' Instructions}
\maketitle

\vspace{-0.5cm}
\input{00_Abstract}
\vspace{-1cm}

\input{01_Introduction}

\input{02_Related_Work}

\input{03_Research_Design}

\input{05_Analysis}

\input{06_Discussion}

\input{08_Conclusions}

\bibliographystyle{unsrt}
\bibliography{bibliography}

\end{document}

%% file: 00_Abstract.tex
\begin{abstract}
[\textbf{Context and motivation}] Ecosystems developed as Open Source Software (OSS) are considered to be highly innovative and reactive to new market trends due to their openness and wide-ranging contributor base. Participation in OSS often implies opening up of the software development process and exposure towards new stakeholders. 
[\textbf{Question/Problem}] 
Firms considering to engage in such an environment should carefully consider potential opportunities and challenges upfront. The openness may lead to higher innovation potential but also to frictional losses for engaged firms. Further, as an ecosystem progresses, power structures and influence on feature selection may fluctuate accordingly. 
[\textbf{Principal ideas/results}] We analyze the Apache Hadoop ecosystem in a quantitative longitudinal case study to investigate changing stakeholder influence and collaboration patterns. Further, we investigate how its innovation and time-to-market evolve at the same time.
[\textbf{Contribution}] Findings show collaborations between and influence shifting among rivaling and non-competing firms. Network analysis proves valuable on how an awareness of past, present and emerging stakeholders, in regards to power structure and collaborations may be created.
Furthermore, the ecosystem's innovation and time-to-market show strong variations among the release history. Indications were also found that these characteristics are influenced by the way how stakeholders collaborate with each other.

\keywords{requirements engineering, stakeholder collaboration, stakeholder influence, open source, software ecosystem, inter-organizational collaboration, open innovation, co-opetition}
\end{abstract}

%% file: 01_Introduction.tex
\section{Introduction}

The paradigm of Open Innovation (OI) encourages firms to look outside for ideas and resources that may further advance their internal innovation capital~\cite{chesbrough2006open}. Conversely, a firm may also find more profitable incentives to open up an intellectual property right (IPR) rather than keeping it closed. For software-intensive firms a common example of such a context is constituted by Open Source Software (OSS) ecosystems~\cite{west2006challenges}~\cite{jansen2009sense}.

The openness implied by OI and an OSS ecosystem makes a firm's formerly closed borders permeable for interaction and influence from new stakeholders, many of which may be unknown to a newly opened-up firm. Entering such an ecosystem affects the way how Requirements Engineering (RE) processes are structured~\cite{linaaker2015requirements}. Traditionally these are centralized, and limited to a defined set of stakeholders. However, in this new open context, RE has moved to become more decentralized and collaborative with an evolving set of stakeholders. This may lead to an increased innovation potential for a firm's technology and product offerings, but also imply frictional losses~\cite{dahlander2008firms}. Conflicting interests and strategies may arise, which may diminish a firms own impact in regards to feature selection and control of product planning~\cite{wnuk2012can}. Further, as an ecosystem evolves, power structures and influence among stakeholders may fluctuate accordingly. This creates a need for firms already engaged or thinking of entering an OSS ecosystem to have an awareness of past and present ecosystem governance constellation in order to be able to adapt their strategies and product planning to upcoming directions of the ecosystem~\cite{jansen2009business}. 

Given this problematization, we were interested in studying how stakeholders' influence and collaboration fluctuate over time in OSS ecosystems. 
Researchers argue that collaboration is core to increase innovation and reduce time-to-market \cite{enkel2009open}. Hence, another goal was to study the evolution of OSS ecosystems' innovation and time-to-market over time. We hypothesize that this could be used as input to firms' planning of contribution and product strategies, which led us to formulate the following research questions:

\begin{enumerate}[RQ1,leftmargin=0.9cm]
\item How are stakeholder influence and collaboration evolving over time? 
\item How are innovation and time-to-market evolving over the same time?
\end{enumerate}

To address these questions, we launched an exploratory and quantitative longitudinal case study of the Apache Hadoop ecosystem, a widely adopted OSS framework for distribution and process parallelization of large data. 

The rest of the paper is structured as follows: Section 2 presents related work. Section 3 describes the case study design and methodology used, limitations and threats to validity are also accounted for. Section 4 presents the analysis and results, which are further discussed in Section 5. Finally, Section 6 concludes the paper.

%% file: 02_Related_Work.tex
\section{Related Work}
Here we present related work to software ecosystems and how its actors (stakeholders) may be analyzed. Further, the fields of stakeholder identification and analysis in RE are presented from an ecosystem and social network perspective.

\subsection{Software Ecosystems}
Multiple definitions of a software ecosystem exists~\cite{manikas2013software}, while we refer to the one by Jansen et al.~\cite{jansen2009sense} - \textit{"A software ecosystem is a set of actors functioning as a unit and interacting with a shared market for software and services, together with relationships among them. These relationships are frequently underpinned by a common technological platform or market and operates through the exchange of information, resources and artifacts."}. The definition may incorporate numerous types of ecosystems in regards to openness~\cite{jansen2012shades}, ranging from proprietary to OSS ecosystems~\cite{manikas2013software}, which in turn contains multiple facets. In this study we will focus on the latter with the Apache Hadoop ecosystem as our case, where the Apache Hadoop project constitutes the technological platform underpinning the relationships between the actors of the Apache Hadoop ecosystem.

An ecosystem may further be seen from three scope levels, as proposed by Jansen et al.~\cite{jansen2009business}. Scope level 1 takes an upper perspective, on the relationships and interactions between ecosystems, for example between the Apache Hadoop and the Apache Spark ecosystems, where the latter's project may be built on top of the former. On scope level 2, one looks inside of the ecosystem, its actors and the relationships between them, which is the focus of this paper when analyzing the Apache Hadoop ecosystem. Lastly, scope level 3 takes the perspective from a single actor and its specific relationships.

Jansen et al.~\cite{jansen2009business} further distinguished between three types of actors: dominators, keystone players, and niche players. Dominators expand and assimilate, often on the expense of other actors. Keystone players are well connected, often with a central role in hubs of actors. They create and contribute value, often beneficial to its surrounding actors. Platform suppliers are typically keystone players. Niche players thrive on the keystone players and strive to distinguish themselves from other niche players. Although other classifications exist~\cite{manikas2013software} ~\cite{jansen2012shades}, we will stick to those defined above.

In the context of OSS ecosystems, a further type of distinction can be made in regards to the Onion model as proposed by Nakakoji et al.~\cite{nakakoji2002evolution}. They distinguished between eight roles ranging the passive user in the outer layer, to the project leader located in the center of the model. For each layer towards the center, influence in the ecosystem increases. Advancement is correlated to increase of contributions and engagement of the user, relating to the concept of meritocracy. 

\subsection{Stakeholder Networks and Interaction in Requirements Engineering}
To know the requirements and constraints of a software, one needs to know who the stakeholders are, hence highlighting the importance of stakeholder identification and analysis in RE~\cite{glinz2007guest}. Knowing which stakeholders are present is however not limited to purposes of requirements elicitation. For firms engaged in OSS ecosystems~\cite{jansen2009sense}~\cite{manikas2013software}, this is important input to their product planning and contribution strategies. Disclosure of differentiating features to competitors, un-synced release cycles, extra patch-work and missed out collaboration opportunities are some possible consequences if the identification and analysis of the ecosystem's stakeholders is not done properly~\cite{wnuk2012can}~\cite{west2006challenges}~\cite{dahlander2008firms}. Most identification methods however refer to the context of traditional software development and lack empirical validation in the context of OSS ecosystems~\cite{pacheco2012systematic}. 

In recent years, the research focus within the field has shifted more towards stakeholder characterization through the use of, e.g., Social Network Analysis (SNA)~\cite{pacheco2012systematic}. It has also become a popular tool in empirical studies of OSS ecosystems, hence highlighting potential application within stakeholder identification.

In regards to traditional software development, Damian et al.~\cite{damian2007collaboration} used SNA to investigate collaboration patterns and the awareness between stakeholders of co-developed requirements in the context of global software development. Lim et al.~\cite{lim2010stakenet} constructed a system based on referrals, where identified stakeholders may recommend others. Concerning RE processes within software ecosystems in general, research is rather limited~\cite{fricker2010requirements} with some exceptions~\cite{knauss2014openness}. Fricker~\cite{fricker2010requirements} proposed that stakeholder relations in software ecosystems may be modeled as requirement value chains \textit{`` \ldots where  requirements emerge from and propagate with inter-stakeholder collaboration''}. Knauss et al.~\cite{knauss2014openness} investigated the IBM CLM ecosystem to find RE challenges and practices used in open-commercial software ecosystems. Distinction is made between a strategic and an emergent requirements flow, where the former regard high level requirements, and how business goals affect the release planning. The latter considers requirements created on an operational level, in a Just-In-Time (JIT) fashion, commonly observed in OSS ecosystems~\cite{ernst2012case}.

In OSS ecosystems specifically, RE practices such as elicitation, prioritization, and selection are usually managed through open forums such as issue trackers or mailinglists. These are also referred to as informalisms as they are used to specify and manage the requirements in an informal manner~\cite{scacchi2002understanding}, usually as a part of a conversation between stakeholders. These informalisms constitute an important source to identify relevant stakeholders. Earlier work includes Duc et al.~\cite{duc2011impact} who applied SNA to map stakeholders in groups of reporters, assignees, and commentators to issues with the goal to investigate the impact of stakeholder collaboration on the resolution time of OSS issues. Crowsten et al.~\cite{crowston2005social} performed SNA on 120 OSS projects to investigate communication patterns in regards to interactions in projects' issue trackers.

Many studies focused on a developer and user level, though some exceptions exist. For example, Martinez-Romeo et al.~\cite{martinez2008using} investigated how a community and a firm collaborates through the development process. Orucevic-Alagic et al.~\cite{orucevic2014network} investigated the influence of stakeholders on each other in the Android project. Texiera et al.~\cite{teixeira2015lessons} explored collaboration between firms in the Openstack ecosystem from a co-opetition perspective showing how firms, despite being competitors, may still collaborate within an ecosystem.

This paper contributes to OSS RE literature by addressing the area of stakeholder identification and analysis in OSS ecosystems by investigating a case on a functional level~\cite{teixeira2015lessons}. Further it adds to the software ecosystem literature and its shallow research of RE~\cite{fricker2010requirements}~\cite{knauss2014openness} and strategic perspectives\cite{manikas2013software} in general.

%% file: 03_Research_Design.tex
\section{Research Design}
We chose the Apache Hadoop project for an embedded case study~\cite{runeson2009guidelines} due to its systematically organized contribution process and its ecosystem composition. Most of the contributors have a corporate affiliation.

To create a longitudinal perspective, issues of the Apache Hadoop's issue tracking and project management tool were analyzed in sets reflecting the release cycles. The analysis was narrowed down to sub releases, spanning from 2.2.0 (released 15/Oct/13) to 2.7.1 (06/Jul/15), thus constituting the units of analysis through the study. Third level releases were aggregated into their parent upper level release.

\begin{figure}[t!]
	\centering
	\includegraphics[width=1.0\linewidth]{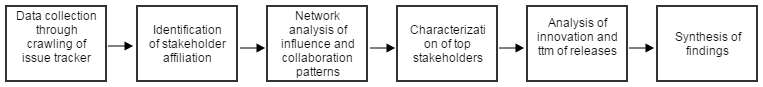}
	\caption{Overview of the case study process}
	\label{fig:analysisProcess}
	\vspace{-0.5cm}
\end{figure}

Issues were furthermore chosen as the main data source as these can tie stakeholders' socio-technical interaction together~\cite{damian2007collaboration}~\cite{duc2011impact}, as well as being connected to a specific release. To determine who collaborated with whom through an issue, patches submitted by each stakeholder were analyzed, a methodology similar to those used in previous studies~\cite{orucevic2014network}~\cite{martinez2008using}. Users who contribute to an issue package their code into a patch and then attach it to the issue in question. After passing a two-step approval process comprising automated tests and manual code reviews, an authorized committer eventually commits the patch to the project's source configuration management (SCM) system.
The overall process of this case study is illustrated in Fig.~\ref{fig:analysisProcess} and further elaborated on below.

\subsection{Data Collection}
The Apache Hadoop project manages its issue data with the issue tracker JIRA. A crawler was implemented to automatically collect, parse, and index the data into a relational database. 

To determine the issue contributors' organizational affiliation, the domain of their email addresses was analyzed. If the affiliation could not be determined directly (e.g., for @apache.org), secondary sources were used such as LinkedIn and Google. The issue contributors' full name functioned as keyword.

\subsection{Analysis Approach and Metrics}
Below we present the methodology and metrics used in the analysis of this paper. Further discussion of metrics in relation to threats to validity is available in section~\ref{sec:threats}.

\begin{wrapfigure}{R}{0.25\textwidth}
\vspace{-1cm}
\centering
\includegraphics[width=0.25\textwidth]{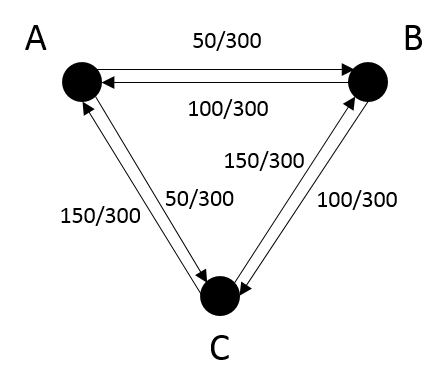}
\caption{\footnotesize Example of a weighted network with three stakeholders.}
\label{fig:NetworkExample}
\vspace{-0.5cm}
\end{wrapfigure}

\vspace{1em}\noindent\textbf{Network Analysis}. Patches attached to issues were used as input to the SNA process. Stakeholders were paired if they submitted a patch to the same issue. Based on stakeholders' affiliation, pairings were aggregated to the organizational level. A directed network was constructed, representing the stakeholders at the organizational level as vertices. Stakeholder collaboration relationships were represented as edges. As suggested by Orucevic-Alagic et al.~\cite{orucevic2014network}, edge weights were calculated to describe the strength of the relationships. Since stakeholders created patches of different size, the relative size of a stakeholder's patch was used for the weighting. We quantified this size as changed lines of code (LOC) per patch.
A simplified example of calculating network weights without organizational aggregation is shown in Fig.~\ref{fig:NetworkExample}. Each of the stakeholders A, B, and C created a patch that was attached to the same issue. A's patch contains 50 LOC. B's patch contains 100 LOC, while C's patch contains 150 LOC. In total, 300 LOC were contributed to the issue. Resulting in the following edge weights: A$\rightarrow$B = 50/300, A$\rightarrow$C = 50/300, B$\rightarrow$C = 100/300, B$\rightarrow$A = 100/300, C$\rightarrow$B = 150/300, and C$\rightarrow$A = 150/300.








The following network metrics were used to measure the influence of stakeholders and the strength of the collaboration relationships among the stakeholders.
\begin{itemize}[nosep]\footnotesize
\item \textit{Out-degree Centrality} is the sum of a all outgoing edges' weights of a stakeholder vertex. Since it calculates the number of collaborations where the stakeholder has contributed, a higher index indicates a higher influence of a stakeholder on its collaborators. It also quantifies the degree of contributions relative to the stakeholder's collaborators.

\item \textit{Betweeness Centrality} counts how often a stakeholder is on a stakeholder collaboration path. A higher index indicates that the stakeholder has a more central position compared to other stakeholders among these collaboration paths.

\item \textit{Closeness Centrality} measures the average relative distance to all other stakeholders in the network based on the shortest paths. A higher index indicates that a stakeholder is well connected and has better possibilities in spreading information in the network, hence a higher influence.

\item \textit{Average Clustering Coefficient} quantifies the degree to which stakeholders tend to form clusters (connected groups). A higher coefficient indicates a higher clustering, e.g., a more densely connected group of stakeholders with a higher degree of collaborations.

\item \textit{Graph Density} is the actual number of stakeholder relationships divided by the possible number of stakeholder relationships. A higher value indicates a better completeness of stakeholder relationships (collaborations) within the network, where 1 is complete and 0 means that no relationships exist.
\end{itemize}

\noindent \textbf{Innovation and Time-To-Market Analysis}. Innovation can be measured through input, output, or process measures \cite{knight2005metrics}. In this study, input and output measures are used to quantify innovation per release. Time-to-market was measured through the release cycle time \cite{griffin1993metrics}.
\begin{itemize}[nosep]\footnotesize
\item \textit{Issues} counts the total number of implemented JIRA tickets per release and comprises the JIRA issue types \emph{feature}, \emph{improvement}, and \emph{bug}. It quantifies the innovation input to the development process. 
\item \textit{Change size} counts the net value of changed lines of code. It quantifies the innovation output of the development process.
\item \textit{Release cycle time} is the amount of time between the start of a release and the end of a release. It indicates the length of a release cycle.
\end{itemize}

\vspace{1em}
\noindent \textbf{Stakeholder Characterization}. 
To complement our quantitative analysis and add further context, we did an qualitative analysis of electronic data available to characterize identified corporate stakeholders. This analysis primarily included their respective websites, press releases, news articles, and blog posts. 


\subsection{Threats to Validity}\label{sec:threats}

Four aspects of validity in regards to a case study are \textit{construct}, \textit{internal} and \textit{external validity}, and \textit{reliability}~\cite{runeson2009guidelines}. 

In regards to \textit{construct validity}, one concern may be definition and interpretation of network metrics. The use of weights to better represent a stakeholder's influence, as suggested by Orucevic-Alagic et al.~\cite{orucevic2014network} was used with the adoption to consider the net of added LOC to further consider the relative size of contributions. A higher number of LOC however does not have to imply increased complexity. We chose to see it as a simplified metric of investment with each LOC representing a cost from stakeholder. Other options could include consideration software metrics such as cyclomatic complexity. Further network metrics, e.g. the eigenvector centrality and the clustering coefficient could offer further facets but was excluded as a design choice.

Furthermore, we focused on input (number of issues) and output (implementation change size) related metrics \cite{knight2005metrics} for operationalizing the innovation per release. Issues is one of many concepts in how requirements may be framed and communicated in OSS RE, hence the term requirement is not always used explicitly~\cite{scacchi2002understanding}. Types of issues varies between OSS ecosystem and type of issue tracker (e.g., JIRA, BugZilla)~\cite{ernst2012case}. In the Apache Hadoop ecosystem we have chosen the types feature, improvement and bug to represent the degree of innovation. We hypothesize that stakeholders engaged in bug fixing, are also involved in the innovation process, even if a new feature and an improvement probably includes a higher degree of novelty in the innovation. Even bugs may actually include requirements-related information not found elsewhere, and also relate to previously defined features with missing information. In future work, weights could be introduced to consider different degrees of innovation in the different issue types.

Release cycle times were used for quantifying the time-to-market as suggested by Griffin~\cite{griffin1993metrics}. Since we solely analyzed releases from the time where the Apache Hadoop ecosystem was already well established, a drawback is that a long requirements analysis ramp up time may not be covered by this measure.

A threat to \textit{internal validity} concerns the observed correlation of how the time-to-market and the innovativeness of a release is influenced by the way how stakeholders collaborate with each other. This needs further replication and validation in future work.

In regards to \textit{external validity}, this is an exploratory single case study. Hence observations need validation and verification in upcoming studies in order for findings to be further generalized. Another limitation concerns that only patches of issues were analyzed, though it has been considered a valid approach in earlier studies~\cite{orucevic2014network}~\cite{martinez2008using}. In future work, consideration should also be taken into account, for example, as this may also be an indicator of influence and collaboration. Further, number of releases in this study was limited due to a complicated release history in the Apache Hadoop project, but also a design choice to give a further qualitative view of each release in a relative fine-grained time-perspective. Future studies should strive to analyze longer periods of time.

Finally, in regards to \textit{reliability} one concern may be the identification of stakeholder affiliation. A contributor could have used the same e-mail but from different roles, e.g., as an individual or for the firm. Further, sources such as LinkedIn may be out of date.

%% file: 05_Analysis.tex
\section{Analysis}\label{sec:analysis}
In this section, we present our results of the quantitative analysis of the Apache Hadoop ecosystem across the six releases R2.2-R2.7.


\subsection{Stakeholders' Characteristics}
Prior to quantitatively analyzing the stakeholder network, we qualitatively analyzed stakeholders' characteristics to gain a better understanding of our studied case.
First, we analyzed how each stakeholder uses the Apache Hadoop platform to support its own business model. We identified the following five user categories:

\begin{itemize}[nosep]\footnotesize
\item \textbf{Infrastructure provider}: sells infrastructure that is based on Apache Hadoop.
\item \textbf{Platform user}: uses Apache Hadoop to store and process data.
\item \textbf{Product provider}: sells packaged Apache Hadoop solutions.
\item \textbf{Product supporter}: Provides Apache Hadoop support without being a product provider.
\item \textbf{Service provider}: Sells Apache Hadoop related services.
\end{itemize}



\noindent Second, we analyzed stakeholders' firm history and strategic business goals to gain a better understanding of their motivation for engaging in the Hadoop ecosystem. We summarize the results of this analysis in the following list:
\begin{itemize}[nosep]\footnotesize
\item \textbf{Wandisco} [Infrastructure provider] entered the Apache Hadoop ecosystem by acquiring AltoStar in 2012. It develops a platform to distribute data over multiple Apache Hadoop clusters.
\item \textbf{Baidu} [Platform user] is a web service company and was founded in 2000. It uses Apache Hadoop for data storage and processing of data.
\item \textbf{eBay} [Platform user] is an E-commerce firm and was founded in 1995. It uses Hadoop for data storage and processing of data.
\item \textbf{Twitter} [Platform user] offers online social networking services and was founded in 2006. It uses Apache Hadoop for data storage and processing of data.
\item \textbf{Xiaomi} [Platform user] is focused on smartphone development. It uses Apache Hadoop for data storage and processing of data.
\item \textbf{Yahoo} [Platform user] is a search engine provider who initiated the Apache Hadoop project in 2005. It uses Apache Hadoop for data storage and processing of data. It spun off Hortonworks in 2011.
\item \textbf{Cloudera} [Product provider] was founded in 2008. It develops its own Apache Hadoop based product \textit{Cloudera Distribution Including Apache Hadoop} (CDH).
\item \textbf{Hortonworks} [Product provider] was spun off by Yahoo in 2011. It develops its own Apache Hadoop based product \textit{Hortonworks Data Platform} (HDP). It collaborates with Microsoft since 2011 to develop \textit{HDP for Windows}. Other partnerships include Redhat, SAP, and Terradata.
\item \textbf{Huawei} [Product provider] offers the Enterprise platform \textit{FusionInsight} based on Apache Hadoop. FusionInsight was first released in 2013.
\item \textbf{Intel} [Product supporter] maintained its own Apache Hadoop distribution that was optimized to their own hardware. It dropped the development in 2014 to support Cloudera by becoming its biggest shareholder and focusing on contributing its features to Cloudera's distribution.
\item \textbf{Altiscale} [Service provider] was founded in 2012. It runs its own infrastructure and offers Apache Hadoop as-a-service via their product \textit{Altiscale Data Cloud}.
\item \textbf{Microsoft} [Service provider] offers Apache Hadoop as a cloud service labeled \textit{HDInsight} through its cloud platform Azure. It maintains a partnership with Hortonworks who develops \textit{HDP for Windows}.
\item \textbf{NTT Data} [Service provider] is a partner with Cloudera and provides support and consulting services for their Apache Hadoop distribution.
\end{itemize}

\noindent Firms that belong to the same user category apply similar business models. Hence, we can identify competing firms based on their categorization. 

\subsection{Stakeholder Collaboration}
\begin{figure*}[t!]
\centering
\includegraphics[scale=0.27]{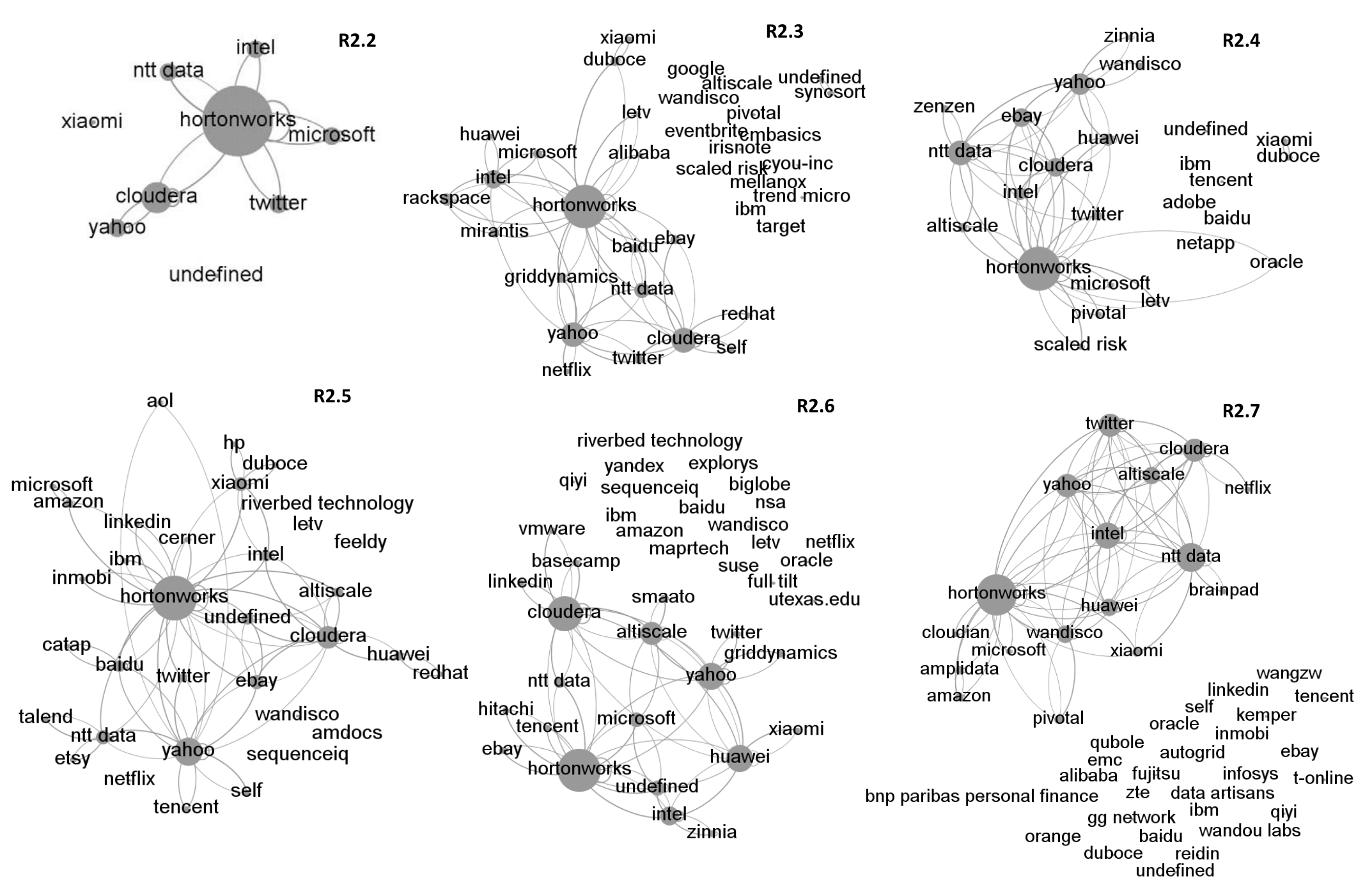}
\caption{Network distribution of releases R2.2-R2.7}
\label{fig:NetworkDistributions}
\end{figure*}

Figure~\ref{fig:NetworkDistributions} shows all stakeholder networks that were generated for the releases R2.2 to R2.7. 
The size of a stakeholder vertex indicates its relative ranking in regards to the outdegree centrality. Table~\ref{tbl:NumberOfVerticesAndEdges} summarizes the number of stakeholders and stakeholder relationships per release. It illustrates that the number of stakeholders and collaboration relationships varies over time. Except for the major increase from R2.2 to R2.3, the network maintains a relatively consistent size, though the number of collaborations are in the interval between 81 to 122 for R2.4 to R2.7.

\begin{table}[]
\vspace{-0.5cm}
\centering
\caption{Number of stakeholder (vertices) and collaboration relationships (edges) per release}
\label{tbl:NumberOfVerticesAndEdges}
\begin{tabular}{|p{3.5cm}|p{1cm}|p{1cm}|p{1cm}|p{1cm}|p{1cm}|p{1cm}|}
\hline
      & R2.2 & R2.3 & R2.4 & R2.5 & R2.6 & R2.7 \\ \hline
Stakeholders & 9    & 35   & 25   & 34   & 38   & 44   \\ \hline
Collaboration relationships & 21   & 97   & 81   & 108  & 96   & 122  \\ \hline
\end{tabular}
\vspace{-0.5cm}
\end{table}

A general observation among the different releases is the existence of one main cluster where a core of stakeholders is present, whilst the remaining stakeholders make temporary appearances. Many stakeholders are not part of these clusters implying that they do not collaborate with other stakeholders at all. The number of those stakeholders shows strong variation among the releases. This could imply that stakeholders implement their own issues, which is further supported by the fact that 65\% of the patches are contributed by the issue reporters themselves. 

\begin{table}[b!]
\vspace{-0.5cm}
\centering
\caption{Average Clustering Coefficient (ACC) and Graph Density (GD) per release.}
\label{tbl:ACCGD}
\begin{tabular}{|p{1.5cm}|p{1cm}|p{1cm}|p{1cm}|p{1cm}|p{1cm}|p{1cm}|}
\hline
    & R2.2  & R2.3  & R2.4  & R2.5  & R2.6  & R2.7  \\ \hline
ACC & 0     & 0.207 & 0.303 & 0.198 & 0.237 & 0.552 \\ \hline
GD  & 0.292 & 0.082 & 0.135 & 0.096 & 0.068 & 0.064 \\ \hline
\end{tabular}
\end{table} 

The visual observation from the networks being weakly connected in general is supported by the Graph Density (GD) as its values are relatively low among all releases (see Table~\ref{tbl:ACCGD}). The values describe that stakeholders had a low number of collaborations in relation to the possible number of collaborations. The Average Clustering Coefficient (ACC) values among all releases (see Table~\ref{tbl:ACCGD}) further indicate that the stakeholders are weakly connected to their direct neighbors in the releases R2.2 - R2.6. This correlates with the observation that there are many unconnected stakeholders and only a few core stakeholders collaborating with each other. The ACC value however indicates a significantly higher number of collaborations for release R2.7. 

\begin{table}[t!]
\centering
\caption{Stakeholder collaborations among the different user categories.}
\label{tbl:collabUserGroups}
\begin{tabular}{|l|p{2cm}|p{1.5cm}|p{1.5cm}|p{1.5cm}|p{1.5cm}|}
\hline
 & Infrastructure provider & Platform user & Product provider & Product supporter & Service provider \\ \hline
Infrastructure provider & 0 & 2 & 4 & 1 & 0 \\ \hline
Platform user & 2 & 24 & 73 & 6 & 14 \\ \hline
Product provider & 4 & 73 & 124 & 23 & 50 \\ \hline
Product supporter & 1 & 6 & 23 & 0 & 3 \\ \hline
Service provider & 0 & 14 & 50 & 3 & 10 \\ \hline
\end{tabular}
\vspace{-0.5cm}
\end{table}

Table~\ref{tbl:collabUserGroups} summarizes stakeholder collaborations among the different user categories. It shows that collaborations took place among all user categories, except between infrastructure providers and service providers. The product providers were the most active and had the highest number of collaborations with other product providers. They also have the highest amount of collaborations with other user categories. These results show that stakeholders with competing (same user category) and non-competing (different user category) business models collaborate within the Apache Hadoop ecosystem.


\subsection{Stakeholder Influence}
To analyze the evolving stakeholder influence over time, we leveraged the three network centrality metrics: outdegree centrality, betweeness centrality, and closeness centrality. 

\begin{figure}[b!]
\vspace{-0.6cm}
  \centering
  \includegraphics[trim=60 350 30 50 ,width=1.1\linewidth]{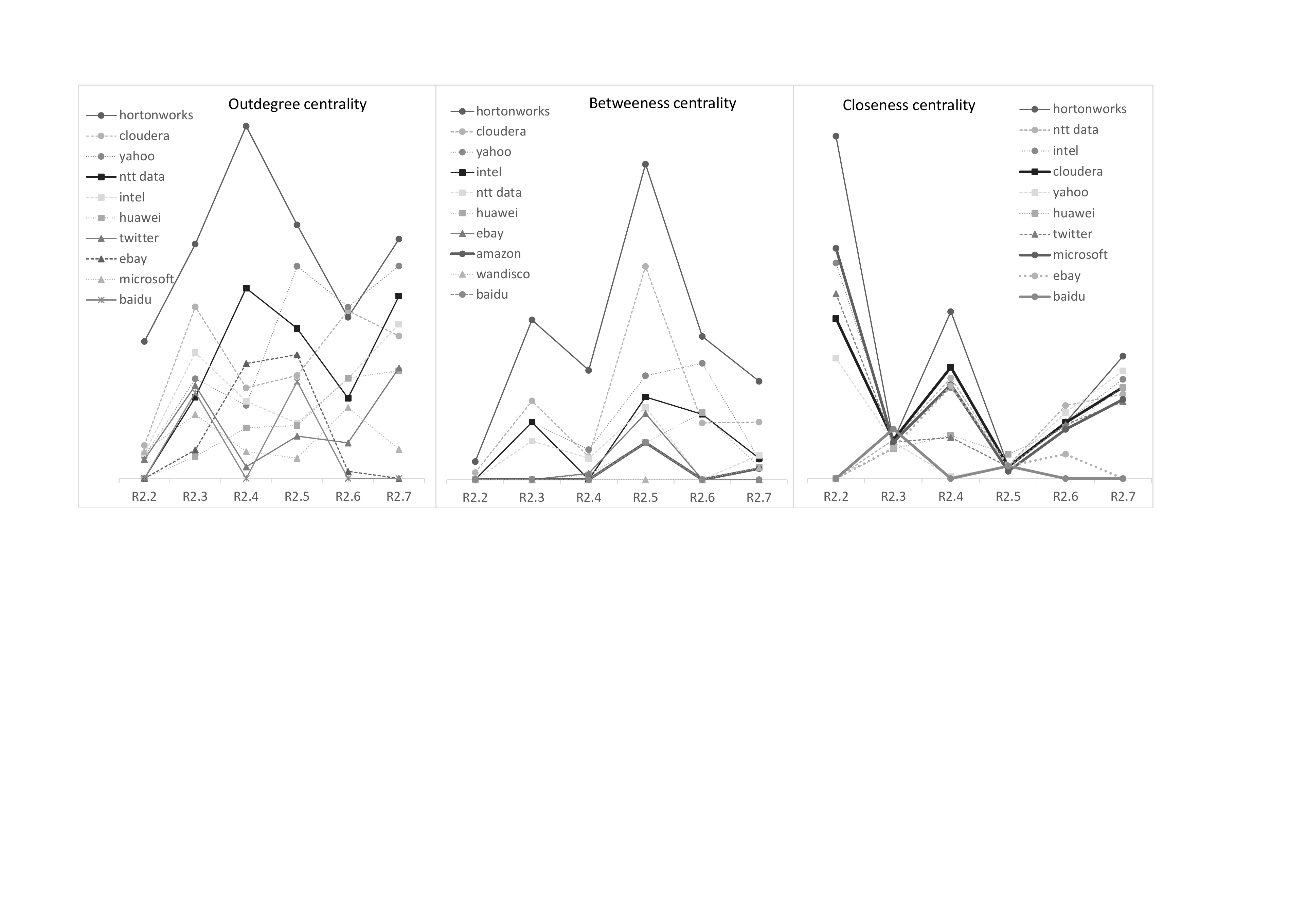}
    \caption{Evolution of stakeholders' outdegree, betweeness, and closeness centrality across the releases R2.2-R2.7}
\label{fig:Graphs}
\end{figure}

The left graph in Fig.~\ref{fig:Graphs} shows the outdegree centrality evolution for the ten stakeholders with the highest outdegree centrality values. These stakeholders are most influential among all Apache Hadoop stakeholders in regards to weighted issue contributions. The graph also shows that the relative outdegree centrality varies over time. To further investigate this evolution, we created a stakeholder ranking per release using the relative outdegree centrality as ranking criteria. This analysis revealed that Hortonworks was most influential in terms of issue contributions. It was five times ranked first and once ranked third (average ranking: 1.3). The other top ranked stakeholders were Cloudera (average ranking: 3.3) and Yahoo (average ranking: 3.3). The stakeholders NTT Data (avg ranking = 4.7) and Intel (average ranking: 4.8) can be considered as intermediate influencing among the top ten outdegree centrality stakeholders. The stakeholders Huawei (average ranking: 8.2), Twitter (average ranking: 8.5), eBay (average ranking: 9.0), Microsoft (average ranking: 9.5), and Baidu (average ranking: 10.2) had the least relative outdegree centrality among the ten stakeholders.





The center graph in Fig.~\ref{fig:Graphs} shows the betweeness centrality evolution of the ten stakeholders with the highest accumulated values. As the metric is based on the number of shortest paths passing through a stakeholder vertex, it indicates a stakeholder's centrality with regards to the possible number of collaborations. The resulting top ten stakeholder list is very similar to the list of stakeholders with the highest outdegree centrality. The top stakeholders are Hortonworks (average ranking: 1), Cloudera (average ranking: 2.7), and Yahoo (average ranking: 3.0). Intel (average ranking: 4.2), NTT Data (average ranking: 4.7), and Huawei (average ranking: 5.3) are influencing among the top ten beweeness centrality stakeholders. eBay (average ranking: 6.7), Amazon (average ranking: 6.7), WANdisco (average ranking: 7.0), and Baidu (average ranking: 7.2), the group of stakeholders with the least betweeness centrality among the top ten stakeholders differs compared to the group of stakeholders with the least outdegree centrality. The stakeholders Twitter and Microsoft were replaced by Amazon and WANdisco.

The right graph in Fig.~\ref{fig:Graphs} shows closeness centrality evolution of the ten stakeholders with the highest accumulated values. A higher degree of closeness centrality indicates higher influence, because of closer collaboration relationships to other stakeholders. The resulting top ten closeness centrality stakeholder list differs compared to the outdegree and betweeness centrality list. Our analysis results do not show a single top stakeholder with the highest closeness centrality. The stakeholders Hortonworks (avgerage ranking: 3.2), NTT Data (average ranking: 4.0), Intel (average ranking: 4.3), Cloudera (average ranking: 4.8), and Yahoo (average ranking: 5.5) had relatively similar closeness rankings among the releases. This is also reflected in Fig.~\ref{fig:Graphs} by very similar curve shapes among the stakeholders. Also the remaining stakeholders with lower closeness centrality values had very similar average rankings: Huawei (average ranking: 7.7), Twitter (average ranking: 8.0), Microsoft (average ranking: 8.3), eBay (average ranking: 9.2), Baidu (average ranking: 9.3).

The results of our analysis also show that the stakeholders with the highest outdegree centrality, betweeness centrality, and closeness centrality were distributed among different stakeholder user categories: 4 platform user, 3 product provider, 2 service provider, and 1 product supporter. However, it is notable that the average ranking differs among these user categories. Product providers had the highest average influence ranking. Platform users and service providers had lower influence ranking. This implies that product providers are the most driving forces of the Apache Hadoop ecosystem.

\subsection{Innovation and Time-To-Market over Time}
\begin{figure}[t!]
\centering
\includegraphics[width=1.0\linewidth]{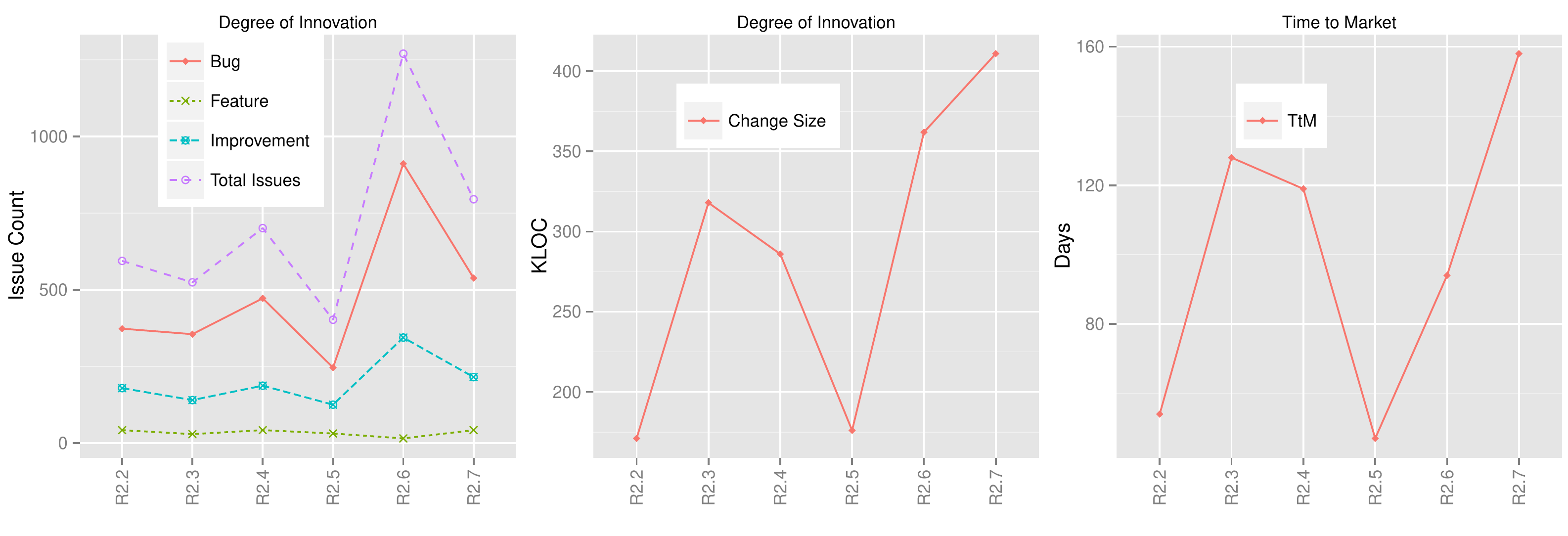}
\caption{Evolution of the degree of innovation over time with respect to implemented JIRA issues and changed lines of code and time to market.}
\label{fig:EvolutionDoI}
\vspace{-0.5cm}
\end{figure}

The evolution of the degree of innovation and time-to-market from release R2.2 to R2.7 is summarized in Figure~\ref{fig:EvolutionDoI} by three consecutive graphs.
The first graph in Fig.~\ref{fig:EvolutionDoI} shows the number of issues that were implemented per release. The illustrated number of issues is broken down into the issue types: bug, improvement, and feature. The number of implemented features (avg: 33.5, med: 37,  std: 9.88) remains steady across all analyzed releases. This is reflected by a relatively low standard deviation. Similarly, the number of implemented improvements (avg: 198.3; med: 183; std: 71.62) remains relatively steady across the releases with one exception. In release R2.6, the double amount of improvement issues was implemented compared to the average of the remaining releases. The number of implemented bugs (avg: 482.5; med: 423;  std: 212.52) features stronger variation among the releases.

The second graph in Fig.~\ref{fig:EvolutionDoI} shows the number of changed lines of code per release. The total number of changed lines of code per release (avg: 287,883.33; med: 302,257; std: 89,334.57) strongly varies across the analyzed releases. Each of the analyzed releases comprises code changes of significant complexity. Even the two releases R2.2 and R2.5, with the lowest change complexity (R2.2: 171 KLOC; R2.5: 176 KLOC), comprised more than 170 KLOC. The remaining releases comprised change complexities of more than 250 KLOC. Further, the graph indicates that the change complexity scatters randomly among the studied releases. A steady trend cannot be determined.

The third graph in Fig.~\ref{fig:EvolutionDoI} depicts the time between the start and the end (time-to-market) of each analyzed release. Analogous to the evolution of the changed lines of code, the time-to-market scatters randomly among the analyzed releases.

%% file: 06_Discussion.tex
\section{Discussion}\label{sec:discussion}

\noindent\textbf{Stakeholder Collaborations (RQ-1)}.
The number of collaborating stakeholders remains on a relatively stable level. However, as indicated by the GD and ACC, the networks are weakly connected in regards to the possible number of collaborations. Only a core set of stakeholders is engaged in most of the collaborations. This may indicate that they have a higher stake in the ecosystem with regards to their product offering and business model, and in turn a keystone behaviour~\cite{jansen2009business}. From a requirements value chain perspective, collaborations translate into partnerships and relationships. This may prove valuable in negotiations about requirements prioritization and how these should be treated when planning releases and road maps~\cite{fricker2010requirements}. The results also show that many stakeholders do not collaborate at all. This is supported by the fact that 65\% of the reported issues are implemented by reporters themselves without any collaboration. This indicates that a lot of independent work was performed in the ecosystem. Reasons for this could be that issues are only of interest for the reporter. It also indicates that the ecosystem is relatively open~\cite{jansen2012shades} in the sense that it is easy for stakeholders to get their own elicited requirements implemented and prioritized, but with the cost of own development efforts.

Another aspect of the collaborations can be inferred from the different user categories. Firms with competing business models collaborate as openly as non-rivaling firms do, as presented in Table~\ref{tbl:collabUserGroups} and reported in earlier studies~\cite{teixeira2015lessons}.
Some of the collaborations may be characterized through the partnerships established between the different stakeholders, as presented in our qualitative analysis of stakeholder characteristics. One of Hortonworks many partnerships include that with Microsoft through the development of their Windows-friendly Apache Hadoop distribution. Cloudera's partnerships include both Intel and NTT Data. None of these partnerships, or among the others identified in this study, occurs within the same user category. Yet still, a substantial part of the ecosystem collaboration occurs outside these special business relationships.  

Independent of business model, all firms work together towards the common goal of advancing the shared platform, much resembling an external joint R\&D pool~\cite{west2006challenges}. As defined through the concept of co-opetition, one motivation could be a joint effort to increase the market share by helping out to create value, and then later diverge and capture value when differentiating in the competition about the customers~\cite{nalebuff1997co}. 
Collaboration could further be limited to commodity parts whereas differentiating parts are kept internal, e.g. leveraged through selective revealing~\cite{henkel2014emergence}. 

\vspace{1em}\noindent\textbf{Stakeholder Influence (RQ-1)}.
Although the distribution of stakeholders' influence fluctuated among the releases, we identified that the group of most influential stakeholders remained very stable. Even the influence ranking within this group did not show high variations. It can be concluded that the development is mainly driven by the stakeholders Hortonworks, Cloudera, NTT Data, Yahoo, and Intel, which may also be referred to as keystone players, and in some cases also niche players relative to each other~\cite{jansen2009business}. Due to this stable evolution, it can be expected that these stakeholders will also be very influential firms in the future. The stakeholder distribution represents multiple user categories, although the product providers Hortonworks and Cloudera tend to be in the top. This may relate to their products being tightly knit with the Apache Hadoop project. In turn, service-providers may use the product-providers' distributions as a basis for their offerings.

Tracking that influence may be useful to identify groups and peers with key positions in order to create traction on certain focus areas for the road map, or to prioritize certain requirements for implementation and release planning~\cite{fricker2010requirements}. Further, it may help to identify emerging stakeholders increasing their contributions and level of engagement~\cite{nakakoji2002evolution}, which may also be reflected in the commercial market. Huawei's increase in outdegree centrality, for example, correlates with the release of their product FusionInsight, which was launched in the beginning of 2013. 


The fact that the network metrics used revealed different top stakeholders, indicates the need of multiple views when analysing the influence. For example, the betweeness centrality Xiaomi, Baidu, and Microsoft in the top compared to the outdegree centrality. This observation indicates that they were involved in more collaboration but produced lower weighted (LOC) contributions relative to their collaborators.



\vspace{1em}\noindent\textbf{Evolution of Ecosystem in Regards to Innovation and Time-To-Market (RQ-2)}.
The analysis results indicate that the number of implemented features does not vary among the analyzed releases. A possible reason for this could be the ecosystem's history. From release R2.2 to R2.5, the project was dominated by one central stakeholder (Hortonworks). Although, additional stakeholders with more influence emerged in release R2.6 and R2.7, Hortonworks remained the dominating contributor, who presumable continued definition and implementation of feature issues. Another potential reason for the lack of variance among features could be the fact that our analysis aggregated all data of third level minor releases to the upper second level releases.

However, our results indicate that the number of implemented improvements show variations among the releases. From release R2.2 to R2.5, the number of implemented improvements per release remained at a steady level. For release R2.6 and R2.7, the number of implemented improvements increased (double the amount). A possible reason for the observed effect could be the fact that other stakeholders with business models get involved in the project to improve the existing ecosystem with respect to their own strategic goals that helps to optimally exploit for their own purpose. The number of implemented bugs varies among all analyzed releases. The high variance of the number of defects could be a side effect of the increased number of improvement issues that potentially imply increase in overall complexity within the ecosystem. Further, the more stakeholders get actively involved in the project to optimize their own business model the more often the ecosystem is potentially used, which may increase the probability to reveal previously undetected defects. 

The analysis results with respect to the evolution of the change size indicate a strong variance among all analyzed releases. Similarly to the change size, the time-to-market measure showed great variance among the analyzed releases. Co-variances of stakeholder collaboration, degree of innovation, and time-to-market measure among the analyzed releases may indicate relationship between these variables. However, to draw this conclusion a detailed regression analysis of multiple ecosystems is required.

\vspace{1em}\noindent\textbf{Implications for Practitioners}.
Even though an ecosystem may have a high population, its governance and project management may still be centered around a small group of stakeholders~\cite{nakakoji2002evolution}, which may further be classified as keystone and in some cases, niche players. Understanding their evolving composition and the influence of these stakeholders may indicate current and possible future directions of the ecosystem~\cite{jansen2009business}. Corporate stakeholders could use this information to better align their open source engagement strategies to their own business goals~\cite{teixeira2015lessons}. It could further provide insights for firms, to what stakeholders' strategic partnerships should be established to improve their strategic influence on the ecosystem regarding, e.g., requirement elicitation, prioritization and release planning~\cite{fricker2010requirements}. Here it is of importance to know how the requirements are communicated throughout the ecosystem, both on a strategic and operational level for a stakeholder to be able to perform the RE processes along with maximized use of its influence~\cite{knauss2014openness}. Potential collaborators may, for example, be characterized with regards to their commitment, area of interest, resource investment and impact~\cite{gonzalez2013understanding}.

The same reasoning also applies for analysis of competitors. Due to the increased openness and decreased distance to competitors implied by joining an ecosystem~\cite{jansen2009business}, it becomes more important and interesting to track what the competitors do~\cite{dahlander2008firms}. Knowing about their existing collaborations, contributions, and interests in specific features offer valuable information about the competitors' strategies and tactics~\cite{teixeira2015lessons}. The methodology used in this study offers an option to such an analysis but needs further research.

Knowledge about stakeholder influence and collaboration patterns may provide important input to stakeholders' strategies. For example, stakeholders may develop strategies on if or when to join an OSS ecosystem, if and how they should adapt their RE processes internally, and how to act together with other stakeholders in an ecosystem using existing practices in OSS RE (e.g.,~\cite{scacchi2002understanding}~\cite{ernst2012case}). This regards both on the strategic and operational level, as requirements may be communicated differently depending on abstraction level, e.g., a focus area for a road map or a feature implementation for an upcoming release~\cite{knauss2014openness}. However, for the operational context in regards to how and when to contribute, further types of performance indicators may be needed. Understanding release cycles and included issues may give an indication of how time-to-market correlates to the complexity and innovativeness of a release. This in turn may help to synchronize a firm's release planning with the ecosystem's, minimizing extra patchwork and missed feature introductions~\cite{wnuk2012can}. Furthermore, it may help a firm planning their own ecosystem contributions and maximize chances for inclusion. In our analysis, we found indications that the time-to-market and the innovativeness of a release is influenced by the way how stakeholders collaborate with each other. Hence, the results could potentially be used as time-to-market and innovativeness predictors for future releases. This however also needs further attention and replication in future research.

%% file: 08_Conclusions.tex
\vspace{-0.3cm}
\section{Conclusions}
\vspace{-0.2cm}

The Apache Hadoop ecosystem is generally weakly connected in regards to collaborations. The network of stakeholders per release consists of a core that is continuously present. A large but fluctuating number of stakeholders work independently. This is emphasized by the fact that a majority of the issues are implemented by the issue reporters themselves. The analysis further shows that the network maintains an even size. One can see that the stakeholders' influence as well as collaborations fluctuate between and among the stakeholders, both competing and non-rivaling. This creates further input and questions to how direct and indirect competitors reason and practically work together, and what strategies are used when sharing knowledge and functionality with each other and the ecosystem. 


In the analysis of stakeholders' influence, a previously proposed methodology was used and advanced to also consider relative size of contributions, and also interactions on an issue level. Further, the methodology demonstrates how an awareness of past, present and emerging stakeholders, in regards to power structure and collaborations may be created. Such an awareness may offer a valuable input to a firm's stakeholder management, and help them to adapt and maintain a sustainable position in an open source ecosystem's governance. Consequently, it may be seen as a pivotal part and enabler for a firm's software development and requirements engineering process, especially considering elicitation, prioritization and release planning for example.


Lastly, we found that innovation and time-to-market of the Apache Hadoop ecosystem strongly varies among the different releases. Indications were also found that these factors are influenced by the way how stakeholders collaborate with each other. 

Future research will focus on what implications stakeholders' influence and collaboration patterns have in an ecosystem. How does it affect time-to-market and innovativeness of a release? How does it affect a stakeholder's impact on feature-selection? How should a firm engaged in an ecosystem adapt and interact in order to maximize its internal innovation process and technology advancement?

\noindent \textbf{Acknowledgments}.
This work was partly funded by the SRC in the SYNERGIES project, Dnr 621-2012-5354, and BMBF grant 01IS14026B.

%% file: main.bbl
\begin{thebibliography}{10}

\bibitem{chesbrough2006open}
Henry~William Chesbrough.
\newblock {\em Open innovation: The new imperative for creating and profiting
  from technology}.
\newblock Harvard Business Press, 2006.

\bibitem{west2006challenges}
Joel West and Scott Gallagher.
\newblock Challenges of open innovation: the paradox of firm investment in
  open-source software.
\newblock {\em R\&d Management}, 36(3):319--331, 2006.

\bibitem{jansen2009sense}
Slinger Jansen, Anthony Finkelstein, and Sjaak Brinkkemper.
\newblock A sense of community: A research agenda for software ecosystems.
\newblock In {\em 31st International Conference on Software Engineering}, pages
  187--190. IEEE, 2009.

\bibitem{linaaker2015requirements}
Johan Lin{\aa}ker, Bj{\"o}rn Regnell, and Hussan Munir.
\newblock Requirements engineering in open innovation: a research agenda.
\newblock In {\em Proceedings of the 2015 International Conference on Software
  and System Process}, pages 208--212. ACM, 2015.

\bibitem{dahlander2008firms}
Linus Dahlander and Mats Magnusson.
\newblock How do firms make use of open source communities?
\newblock {\em Long Range Planning}, 41(6):629--649, 2008.

\bibitem{wnuk2012can}
Krzysztof Wnuk, Dietmar Pfahl, David Callele, and Even-Andr{\'e} Karlsson.
\newblock How can open source software development help requirements management
  gain the potential of open innovation: an exploratory study.
\newblock In {\em Proceedings of the ACM-IEEE international symposium on
  Empirical software engineering and measurement}, pages 271--280. ACM, 2012.

\bibitem{jansen2009business}
Slinger Jansen, Sjaak Brinkkemper, and Anthony Finkelstein.
\newblock Business network management as a survival strategy: A tale of two
  software ecosystems.
\newblock {\em Proccedings of the 1st International Workshop on Software
  Ecosystems}, pages 34--48, 2009.

\bibitem{enkel2009open}
Ellen Enkel, Oliver Gassmann, and Henry Chesbrough.
\newblock Open r\&d and open innovation: exploring the phenomenon.
\newblock {\em R\&d Management}, 39(4):311--316, 2009.

\bibitem{manikas2013software}
Konstantinos Manikas and Klaus~Marius Hansen.
\newblock Software ecosystems--a systematic literature review.
\newblock {\em Journal of Systems and Software}, 86(5):1294--1306, 2013.

\bibitem{jansen2012shades}
Slinger Jansen, Sjaak Brinkkemper, Jurriaan Souer, and Lutzen Luinenburg.
\newblock Shades of gray: Opening up a software producing organization with the
  open software enterprise model.
\newblock {\em Journal of Systems and Software}, 85(7):1495--1510, 2012.

\bibitem{nakakoji2002evolution}
Kumiyo Nakakoji, Yasuhiro Yamamoto, Yoshiyuki Nishinaka, Kouichi Kishida, and
  Yunwen Ye.
\newblock Evolution patterns of open-source software systems and communities.
\newblock In {\em Proceedings of the international workshop on Principles of
  software evolution}, pages 76--85. ACM, 2002.

\bibitem{glinz2007guest}
Martin Glinz and Roel~J Wieringa.
\newblock Guest editors' introduction: Stakeholders in requirements
  engineering.
\newblock {\em IEEE Software}, 24(2):18--20, 2007.

\bibitem{pacheco2012systematic}
Carla Pacheco and Ivan Garcia.
\newblock A systematic literature review of stakeholder identification methods
  in requirements elicitation.
\newblock {\em Journal of Systems and Software}, 85(9):2171--2181, 2012.

\bibitem{damian2007collaboration}
Daniela Damian, Sabrina Marczak, and Irwin Kwan.
\newblock Collaboration patterns and the impact of distance on awareness in
  requirements-centred social networks.
\newblock In {\em 15th IEEE International Requirements Engineering Conference},
  pages 59--68. IEEE, 2007.

\bibitem{lim2010stakenet}
Soo~Ling Lim, Daniele Quercia, and Anthony Finkelstein.
\newblock Stakenet: using social networks to analyse the stakeholders of
  large-scale software projects.
\newblock In {\em Proceedings of the 32nd ACM/IEEE International Conference on
  Software Engineering}, pages 295--304. ACM, 2010.

\bibitem{fricker2010requirements}
Samuel Fricker.
\newblock Requirements value chains: Stakeholder management and requirements
  engineering in software ecosystems.
\newblock In {\em Requirements Engineering: Foundation for Software Quality},
  pages 60--66. Springer, 2010.

\bibitem{knauss2014openness}
Eric Knauss, Daniela Damian, Alessia Knauss, and Arber Borici.
\newblock Openness and requirements: Opportunities and tradeoffs in software
  ecosystems.
\newblock In {\em IEEE 22nd International Requirements Engineering Conference
  (RE)}, pages 213--222. IEEE, 2014.

\bibitem{ernst2012case}
Neil Ernst and Gail~C Murphy.
\newblock Case studies in just-in-time requirements analysis.
\newblock In {\em IEEE Second International Workshop on Empirical Requirements
  Engineering}, pages 25--32. IEEE, 2012.

\bibitem{scacchi2002understanding}
Walt Scacchi.
\newblock Understanding the requirements for developing open source software
  systems.
\newblock In {\em Software, IEE Proceedings-}, volume 149, pages 24--39. IET,
  2002.

\bibitem{duc2011impact}
Anh~Nguyen Duc, Daniela~S Cruzes, Claudia Ayala, and Reidar Conradi.
\newblock Impact of stakeholder type and collaboration on issue resolution time
  in oss projects.
\newblock In {\em Open Source Systems: Grounding Research}, pages 1--16.
  Springer, 2011.

\bibitem{crowston2005social}
Kevin Crowston and James Howison.
\newblock The social structure of free and open source software development.
\newblock {\em First Monday}, 10(2), 2005.

\bibitem{martinez2008using}
Juan Martinez-Romo, Gregorio Robles, Jesus~M Gonzalez-Barahona, and Miguel
  Ortu{\~n}o-Perez.
\newblock Using social network analysis techniques to study collaboration
  between a floss community and a company.
\newblock In {\em Open Source Development, Communities and Quality}, pages
  171--186. Springer, 2008.

\bibitem{orucevic2014network}
Alma Orucevic-Alagic and Martin H{\"o}st.
\newblock Network analysis of a large scale open source project.
\newblock In {\em 40th EUROMICRO Conference on Software Engineering and
  Advanced Applications}, pages 25--29, Verona, Italy, 2014. IEEE.

\bibitem{teixeira2015lessons}
Jose Teixeira, Gregorio Robles, and Jes{\'u}s~M Gonz{\'a}lez-Barahona.
\newblock Lessons learned from applying social network analysis on an
  industrial free/libre/open source software ecosystem.
\newblock {\em Journal of Internet Services and Applications}, 6(1):1--27,
  2015.

\bibitem{runeson2009guidelines}
Per Runeson and Martin H{\"o}st.
\newblock Guidelines for conducting and reporting case study research in
  software engineering.
\newblock {\em Empirical software engineering}, 14(2):131--164, 2009.

\bibitem{knight2005metrics}
Dan Knight, Robert~M Randall, Amy Muller, Liisa V{\"a}likangas, and Paul
  Merlyn.
\newblock Metrics for innovation: guidelines for developing a customized suite
  of innovation metrics.
\newblock {\em Strategy \& Leadership}, 33(1):37--45, 2005.

\bibitem{griffin1993metrics}
Abbie Griffin.
\newblock Metrics for measuring product development cycle time.
\newblock {\em Journal of product innovation management}, 10(2):112--125, 1993.

\bibitem{nalebuff1997co}
Barry~J Nalebuff and Adam~M Brandenburger.
\newblock Co-opetition: Competitive and cooperative business strategies for the
  digital economy.
\newblock {\em Strategy \& leadership}, 25(6):28--33, 1997.

\bibitem{henkel2014emergence}
Joachim Henkel, Simone Sch{\"o}berl, and Oliver Alexy.
\newblock The emergence of openness: How and why firms adopt selective
  revealing in open innovation.
\newblock {\em Research Policy}, 43(5):879--890, 2014.

\bibitem{gonzalez2013understanding}
Jesus~M Gonzalez-Barahona, Daniel Izquierdo-Cortazar, Stefano Maffulli, and
  Gregorio Robles.
\newblock Understanding how companies interact with free software communities.
\newblock {\em IEEE software}, (5):38--45, 2013.

\end{thebibliography}
